\documentclass[sigconf]{acmart}

\acmConference[]{}{}{}

\usepackage[utf8]{inputenc} 
\usepackage[T1]{fontenc}    
\usepackage{hyperref}       
\usepackage{url}            
\usepackage{booktabs}       
\usepackage{amsfonts}       
\usepackage{nicefrac}       
\usepackage{microtype}      
\usepackage{caption}
\usepackage{subcaption}
\usepackage{graphicx}
\usepackage{booktabs}
\usepackage{xcolor}
\usepackage{listings}
\usepackage{backnaur}
\usepackage{stmaryrd}

\definecolor{codegreen}{rgb}{0,0.6,0}
\definecolor{codegray}{rgb}{0.5,0.5,0.5}
\definecolor{backcolour}{rgb}{0.95,0.95,0.92}

\lstdefinestyle{mystyle}{
    commentstyle=\color{codegreen},
    keywordstyle=\color{blue},
    numberstyle=\tiny\color{codegray},
    stringstyle=\color{red},
    basicstyle=\ttfamily\footnotesize,
    breakatwhitespace=false,         
    captionpos=b,                    
    keepspaces=true,                 
    numbers=left,                    
    numbersep=5pt,                  
    showspaces=false,                
    showstringspaces=false,
    showtabs=false,                  
    tabsize=1
}

\usepackage{booktabs} 
\usepackage{xcolor,colortbl}
\definecolor{Gray}{gray}{0.9}
\definecolor{Green}{RGB}{89,163,76}
\usepackage{fontawesome}
\usepackage{url}
\usepackage{todonotes}
\usepackage{balance}
\usepackage{framed}

\newenvironment{bnfsplit}[1][0.7\textwidth]
 {\minipage[t]{#1}$}
 {$\endminipage}

\title{Mining Idioms in the Wild}
\author{Aishwarya Sivaraman}
 \authornote{Work performed as an intern at Facebook, Menlo Park.}
 \affiliation{%
     \institution{UCLA}
     \country{USA}
}

\author{Rui Abreu}
\authornote{Work performed while on sabbatical at Facebook, Menlo Park.}
\affiliation{%
     \institution{U.Porto/INESC-ID}
     \country{Portugal}
}

\author{Andrew Scott, Tobi Akomolede, Satish Chandra}
\affiliation{%
     \institution{Facebook}
     \country{USA}
}

\setcopyright{none}
\renewcommand\footnotetextcopyrightpermission[1]{} 
\begin{document}

\begin{abstract}

Existing code repositories contain numerous instances of code
patterns that are idiomatic ways of accomplishing a particular programming task. Sometimes, the programming language in use supports specific operators or APIs that can express the same idiomatic imperative code much more succinctly. However, those code patterns linger in repositories because the developers may be unaware of the new APIs, or just have not got around to it.  Detection of idiomatic code can also point to the \emph{need} for new APIs. 

We share our experiences in mine idiomatic patterns from the Hack repo at Facebook.  We found that existing techniques either cannot identify meaningful patterns from syntax trees or require test-suite-based dynamic analysis to incorporate semantic properties to mine useful patterns. The key insight of the approach proposed in this paper --- \emph{Jezero} --- is that semantic idioms from a large codebase can be learned from \emph{canonicalized} dataflow trees. We propose a scalable, lightweight static analysis-based approach to construct such a tree that is well suited to mine semantic idioms using nonparametric Bayesian methods.  

Our experiments with \emph{Jezero} on Hack code shows a clear advantage of adding canonicalized dataflow information to ASTs: \emph{Jezero} was significantly more effective than a baseline that did not have the dataflow augmentation in being able to effectively find refactoring opportunities from unannotated legacy code.


\end{abstract}

\maketitle

\section{Introduction}
\label{sec:introduction}

An idiom is a syntactic fragment that recurs frequently across software projects. 
Idiomatic code is usually the most natural way to express a certain computation,
which explains its frequent recurrence in code.  An idiomatic imperative code fragment often has a single semantic purpose that, in principle, can be replaced with API calls or functional operators.

To illustrate the motivation for this work, consider the imperative code examples in the Hack programming language\footnote{Hack is a programming language for the HipHop Virtual Machine, created by Facebook as a dialect of PHP: \url{https://hacklang.org/}.} 
in Figure~\ref{fig:motivation_idiom_examples}, (a), (c) and (e). These examples---adapted from the codebase at Facebook---loop over a vector and accumulate some value in the loop body. To capture this idiom, Hack supports a more functional
\texttt{Vec$\backslash$map\_with\_key} API, and we do find instances where a developer
refactored code to replace a loop with a map call; for instance, see (b), (d) and (f).  This kind of refactoring is of course not unique to Hack; examples in other programming languages abound, such as using LINQ APIs in C\#~\cite{allamanis2018mining} or Python's list comprehensions. 

Why do imperative idioms continue to linger in code?  This can be attributed to several reasons: (1) developers being unaware of the API that can replace the imperative code, or (2) a new API construct being introduced and imperative locations not being updated consistently to use 
this construct, or (3) an API that would simplify this idiomatic pattern has not been included in the language yet.  Identifying such idiomatic patterns and replacing the idiomatic imperative code with corresponding
API calls or operators can help in maintainability and comprehensibility of the codebase. Additionally, identification of common idioms may provide data-driven guidance to language developers for new language constructs (this purpose is outside this paper's scope).

\begin{figure*}[t!]
\centering

\begin{subfigure}[b]{\textwidth}
     \begin{subfigure}[b]{0.4\textwidth}
\begin{lstlisting}[language=PHP]
$output = '';
foreach ($outputs as $key => $value) {
  $output .= self::getOutput(
    $key, $value,
    $task, $pair, ); }
\end{lstlisting}
                    \caption{Example 1 --- Imperative Version}\label{fig:motivation_eg_1}
                 \end{subfigure}
     \hspace{2cm}
          \begin{subfigure}[b]{0.4\textwidth}
\begin{lstlisting}[language=PHP]
$output .= Str\join(
    Vec\map_with_key(
        $outputs,
        ($key, $value) ==> {
            return self::getOutput(
                $key, $value,
                $task, $pair,); }, ), '',);
\end{lstlisting}
            \caption{Example 1 --- API Version}\label{fig:motivation_idiom_eg_1}
         \end{subfigure}
\end{subfigure}
     \begin{subfigure}[b]{\textwidth}
     \begin{subfigure}[b]{0.4\textwidth}

\begin{lstlisting}[language=PHP]
$call_stack_nodes = vec[];
foreach ($identifier_to_id as $identifier => $id) {
 $call_stack_nodes[] = shape(
   'id' => $id,
   'vertex' => $nodes[$id]
   'tally' => $identifier_to_count[$identifier],
   'fraction' => 
     $this->fraction($identifier_to_count[$identifier], 
                  $total_count), ); }
\end{lstlisting}
                    \caption{Example 2 --- Imperative Version}\label{fig:motivation_eg_2}
                 \end{subfigure}
     \hspace{2cm}
          \begin{subfigure}[b]{0.4\textwidth}
\begin{lstlisting}[language=PHP]
$call_stack_nodes = Vec\map_with_key(
    $identifier_to_id,
    ($identifier, $id) ==> shape(
        'id' => $id,
        'vertex' => $nodes[$id],
        'tally' => $identifier_to_count[$identifier]
        'fraction' => 
            $this->fraction($identifier_to_count[$identifier], 
             $total_count), ); );
\end{lstlisting}
            \caption{Example 2 --- API Version}\label{fig:motivation_idiom_eg_2}
         \end{subfigure}
\end{subfigure}
\begin{subfigure}[b]{\textwidth}
     \begin{subfigure}[b]{0.4\textwidth}
\begin{lstlisting}[language=PHP]
$versions = vec[];
foreach ($ids as $key => $value) {
  $versions[] = tuple($value, $names[$key]); }
\end{lstlisting}
                    \caption{Example 3 --- Imperative Version}\label{fig:motivation_eg_3}
                 \end{subfigure}
     \hspace{2cm}
          \begin{subfigure}[b]{0.4\textwidth}
\begin{lstlisting}[language=PHP]
$versions = Vec\map_with_key(
    $ids,
    ($key, $value) ==> tuple($value, $names[$key]); );
\end{lstlisting}
            \caption{Example 3 --- API Version}\label{fig:motivation_idiom_eg_3}
         \end{subfigure}
\end{subfigure}
        \caption{Examples of idiomatic imperative code and its corresponding version rewritten with map API.}
        \label{fig:motivation_idiom_examples}
\end{figure*}

\subsection{Finding missed refactoring opportunities}

Suppose we want a tool that looks at past instances of refactorings and then identifies additional opportunities of similar refactorings either in existing or new code. For example, the code in Figure~\ref{fig:motivation_eg_1}, ~\ref{fig:motivation_eg_2}, and ~\ref{fig:motivation_eg_3} was replaced, respectively,
with the code in Figure~\ref{fig:motivation_idiom_eg_1},~\ref{fig:motivation_idiom_eg_2}, and ~\ref{fig:motivation_idiom_eg_3}, where each of the examples was rewritten using the map API.  We want the tool to learn a general pattern which when exists in the code, begs for
refactoring.\footnote{An even more powerful tool would also suggest the refactored code that is drop-in replacement for the existing code, but the design of semantically correct code transformation is a separate hard problem, outside the scope of this work.}

At first blush, this may look like a pattern matching or clone detection problem, where a code fragment that is a candidate for refactoring might be a clone of code that \textit{was} refactored in the past to introduce an API call.  Another candidate approach might aim to extract generalizable code transformations from a small set of specific examples of transformations~\cite{bader2019getafix,meng2013lase,miltner2019fly,polozov2015flashmeta,rolim2017learning,rolim2018learning,gao2020feedback}.

However, identifying the pattern for \texttt{Vec$\backslash$map\_with\_key} API---let alone the transformation---from the examples in Figure~\ref{fig:motivation_idiom_examples} is non-trivial for the following reasons:
\begin{enumerate}
    \item Code that maps values to an accumulator may have different types. For example, in Figure~\ref{fig:motivation_eg_1} we have string accumulation whereas in Figure~\ref{fig:motivation_eg_2} we have a vector accumulation. Therefore, any tool that can identify this pattern would need to identify the semantic pattern that each of the example in Figure~\ref{fig:motivation_idiom_examples} is accumulating to a collection variable. 
    \item Code may be interleaved with other code, as in Figure~\ref{fig:motivation_eg_2}. In this example, although the accumulation is to a vector variable, they have additional code that operates on functions and variables other than the key and value. Additionally, code in Figure~\ref{fig:motivation_eg_1} interleaves iteration and string concatenation. 
\end{enumerate}
Therefore, a naive syntactic-pattern-based approach would fail to identify common patterns
that matches all examples in Figure~\ref{fig:motivation_idiom_examples}.
A different potential solution is to define a domain-specific language 
and let developers handcraft custom rules to identify semantic patterns. This requires significant manual 
effort, and less experienced programmers might not know how to generalize these patterns. 

We turn to the statistical pattern mining work by Allamanis and Sutton~\cite{allamanis2014mining}, which has shown the possibility of finding patterns in abstract syntax trees (ASTs) that correspond to idioms based on
their frequency of occurrence.  They proposed to use probabilistic tree substitution grammars (pTSG), a non-parametric Bayesian machine learning technique to find idioms (we give an overview of this technique in Section 
~\ref{sec:stats}).  While this is an exciting idea, in practice this does not work very well out of the box, 
as Allamanis et al. report in their follow-up work~\cite{allamanis2018mining} (and as we found as well, see Section~\ref{sec:evaluation}.)  Because purely syntactic idioms are oblivious to semantics, they capture only shallow patterns that are not useful for our end purpose.

In subsequent work, Allamanis \textit{et al.}~\cite{allamanis2018mining} propose to use ASTs augmented with variable mutability and function purity information (we give an overview in Section~\ref{sec:prior_idiom}.) They found that this worked well for refactoring loops with functional constructs in LINQ.

Unfortunately, we found practical issues with the enhancement in \cite{allamanis2018mining} when it comes to our setting:  (1) 
the technique in \cite{allamanis2018mining} requires manual annotations or dynamic analysis to infer those, neither of which were an option for us; (2) it can only match patterns with exact lexical order of appearance of variables, and (3) it cannot detect patterns interleaved with other code. Based on our pencil and paper simulation with known mutability and purity information, their approach fails to learn a pattern that matches the examples in Figure~\ref{fig:motivation_idiom_examples}. We discuss this limitation further in Section~\ref{sec:prior_idiom}.

\begin{figure*}[ht]
     \begin{subfigure}[b]{0.6\columnwidth}
         \centering
     \includegraphics[scale=0.25]{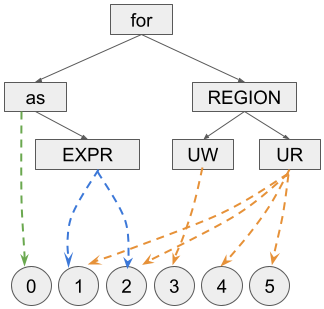}
     \caption{Example 1}\label{fig:semantic_motivation_eg_1}
     \end{subfigure}
     \begin{subfigure}[b]{0.6\columnwidth}
         \centering
        \includegraphics[scale=0.25]{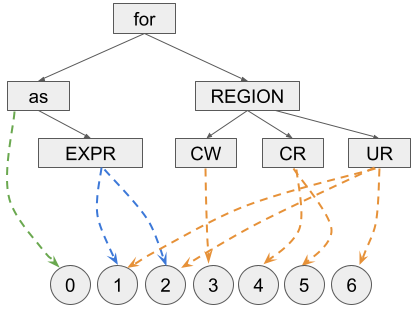}
         \caption{Example 2}
         \label{fig:semantic_motivation_eg_4}
     \end{subfigure}
     \begin{subfigure}[b]{0.6\columnwidth}
        \centering
        \includegraphics[scale=0.25]{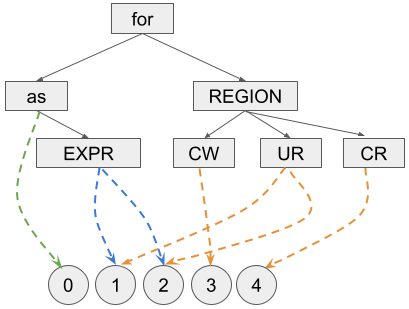}
        \caption{Example 3}
         \label{fig:semantic_motivation_eg_3}
     \end{subfigure}
        \vspace{-0.4cm}
        \caption{Coiled ASTs for the imperative code in Figure~\ref{fig:motivation_idiom_examples}: the basis for computing semantic idioms in Allamanis \textit{et al.}~\cite{allamanis2018mining}.
        The variables map to references illustrated in circles.}
        \label{fig:semantic_loops_motivation_examples}
\end{figure*}

We propose \emph{Jezero}, an approach that works around the practical complexities of the work of Allamanis 
\textit{et al.}~\cite{allamanis2018mining}.  Our approach adds semantic information
to ASTs in a different way: approximate dataflow 
information represented as an extension of the AST. \emph{Jezero} automatically learns semantic patterns from a large codebase over tree structures--dataflow augmented ASTs--that
are generated leveraging a cheap, syntactic dataflow analysis. Our key insight is that semantic 
patterns can be captured as \textit{canonical} dataflow trees. This observation is inspired by 
the seminal Programmer's Apprentice paper~\cite{rich1988programmer} idea that high-level concepts can be
identified as dataflow patterns. In fact, recent works in the area of code search~\cite{premtoon2020semantic},
code clone detection~\cite{wang2020detecting}, and refactoring~\cite{ meng2015does} also use this insight and use dataflow analysis to identify semantically similar code.

For instance, a desirable dataflow pattern that summarizes the examples in 
Figure~\ref{fig:motivation_idiom_examples} is the following: \texttt{foreach} 
contains a \texttt{datawrite} to a \texttt{collection} variable with 
\texttt{dataread} from first and second \texttt{primitive} variables and the first 
\texttt{collection} variable in the order of their definition. Concretely, to learn 
such a pattern, given a set of methods, we collect approximate dataflow information 
from their abstract syntax trees and construct a dataset of canonicalized trees as 
described in Section~\ref{sec:dataflowtrees}.  We then mine for dataflow patterns 
using a non-parametric Bayesian ML technique (see 
Section~\ref{sec:stats})~\cite{allamanis2014mining}, to which, our representation 
looks just like any other tree.  Figure~\ref{fig:top1vecmapwithkey} shows the 
tree pattern that \emph{Jezero} mined and that matches all three examples of 
Figure~\ref{fig:motivation_idiom_examples}.\footnote{Due to the statistical process, the patterns mined sometimes may not cover all relevant aspects of the desirable pattern.}

The process of using \emph{Jezero} involves the following steps: (1) point
\emph{Jezero} at a corpus that is likely to contain instances of the idiomatic pattern we want to uncover; (2) let \emph{Jezero} mine the patterns and come up with a most suitable one(s) using its ranking heuristics; and (3) use \emph{Jezero} further to point out locations in the code where similar refactoring can be carried out.  

\subsection{Contributions}
This paper makes the following contributions:
\begin{itemize}

\item We present a new canonicalized tree representation based on inexpensive dataflow to mine semantic idioms. The approach takes as input a code corpus, generates a tree for each method augmented with dataflow information, and
similar to~\cite{allamanis2014mining}, uses Bayesian learning methods to mine idiomatic patterns.  Our tree
representation overcomes the
practical problems in adopting 
the closest previous work~\cite{allamanis2018mining}.

\item We present \emph{Jezero} a tool that implements both idiom learning and identification of refactoring
opportunities that works at the scale of Facebook code base.
    
\item We evaluate \emph{Jezero} on the task of mining idioms for loopy map/filter code. The mining is done over $1347$ (refactoring) instances per API taken from Facebook's Hack codebase. Each instance contains two commit versions, one version with the imperative code and the other with the code refactored to use a functional operator. 
On an evaluation set, we found
\emph{Jezero}’s F1 score to be
0.57, significantly better than
a baseline technique without the dataflow
enhancement.

    
\item We also evaluated \emph{Jezero} for identifying new, hitherto unknown opportunities for refactoring code to introduce APIs. Using the top-ranked idioms, we then found $807$ matches by matching these idioms against 
the Facebook code base containing $13770$ Hack methods; the average precision of finding real opportunities was $0.60$. The baseline, without dataflow, matched a mere 23 locations.
\end{itemize}

To our knowledge, \textbf{\emph{Jezero} is unique in its ability to find refactoring opportunities from legacy code, based on purely unsupervised learning, and without requiring annotations or dynamic analysis.}  Moreover, we expect the ideas in \emph{Jezero} to carry over to other languages such as Python, which over time have provided more succinct ways to express idiomatic code.
%
%
\section{Background on Mining Idioms}
\label{sec:prior_idiom}
Allamanis and Sutton~\cite{allamanis2014mining} have addressed the 
problem of idiom mining as an unsupervised learning problem and proposed a probabilistic tree substitution grammar (pTSG) inference to mine 
idiomatic patterns. In this work, they mine syntactic idioms from ASTs; 
however, in their following work~\cite{allamanis2018mining} they show that syntactic idioms tend to capture shallow, uninterpretable patterns and fail to capture widely used idioms. Data sparsity and extreme code variability are cited as the 
reasons for shallow idioms. Therefore, to mine interesting idioms and to avoid sparsity, the authors introduce semantic idioms. Semantic idioms improve upon syntactic idioms through a \emph{coiling} process~\cite{allamanis2018mining}. Coiling is a graph transformation that augments standard ASTs with semantic information to yield coiled ASTs (CASTs). These CASTs are then mined using probabilistic tree substitution grammars (pTSG), a machine learning technique for finding salient (and not just frequent) patterns~\cite{cohn2010inducing}. They infer semantic properties such as variable mutability and function purity using a testing-based analysis. For the libraries that do not have test suites, the authors manually annotate with the required properties. The lower path in Fig~\ref{fig:overview} shows the overall process.
 
Using the semantic information, the coiling rewrite phase augments the nodes with variable mutability and distinguishes collections from other variables. The pruning phase retains only subtrees rooted at loop headers and abstracts expressions and control-free statement sequences to \emph{regions} to reduce sparsity. Specifically, they abstract loop expressions into a single \texttt{EXPR} node, labeled with variable references. Additionally, they use \texttt{REGION} nodes to capture code blocks that do not contain any control statements. These nodes encode the purity of variables used in the code block. The purity node types include read (R), write (W), and read-write (RW), and these nodes further differentiate between primitive (prefixed by U) and collection (prefixed by C) variables. Note that region nodes in this work only consider variable mutability, i.e., whether variables, being it collection or primitive, are read from or written to. While this representation effectively captures a class of semantic idioms, it is not sufficient to capture refactoring idioms that require additional flow information between variables. Figure~\ref{fig:semantic_loops_motivation_examples} shows the CASTs for the examples in Figure~\ref{fig:motivation_idiom_examples}. 

Despite the effectiveness of the proposed methods in identifying semantic loop 
idioms, they suffer from limitations that prohibit direct application for 
identifying code patterns as in Figure~\ref{fig:motivation_idiom_examples} --- 
which do occur in realistic and large codebases. Specifically,
\begin{enumerate}
    \item To construct \texttt{REGION} nodes, prior work relied on manual annotations or testing-based analysis. Both these efforts are expensive and might not be available for all codebases. Certainly, this is not available in legacy codebases. 
    \item The augmented trees contain variable references which are numbered based on their lexicographical ordering. Hence, two loops with the same semantic concept but with a different number of variable references will have different patterns. Looking at Figure~\ref{fig:semantic_loops_motivation_examples}, at this level of abstraction, it is not clear that these trees are about the same idiom. 
    \item Further, due to the lexicographical ordering, loops with the same concept but with additional interleaved code statements will most likely have different patterns.
\end{enumerate}
Following these limitations, despite the code in Figure~\ref{fig:motivation_idiom_examples}
being arguably about very similar constructs, Allamanis \textit{et al.}'s 
approach~\cite{allamanis2018mining}
would fail to consider the examples as being part of the same idiom. 
The desired idioms, shown 
in Figure~\ref{fig:semantic_loops_motivation_examples}, albeit similar, are sufficiently 
distinct (e.g., ordering of variables) to be considered the same.


While this is the state-of-the-art in idiom mining, the fact that it requires 
dynamic analysis makes it impractical to be used in our codebase. As such, 
in Section~\ref{sec:evaluation}, we 
will instead compare our approach with the AST-based tree representation for
idiom mining proposed by Allamanis \textit{et al.}~\cite{allamanis2014mining}.
\section{Proposed Approach: \emph{Jezero}}
\label{sec:aug_trees}
\begin{figure}[h!]
    \centering
    \includegraphics[scale=0.2]{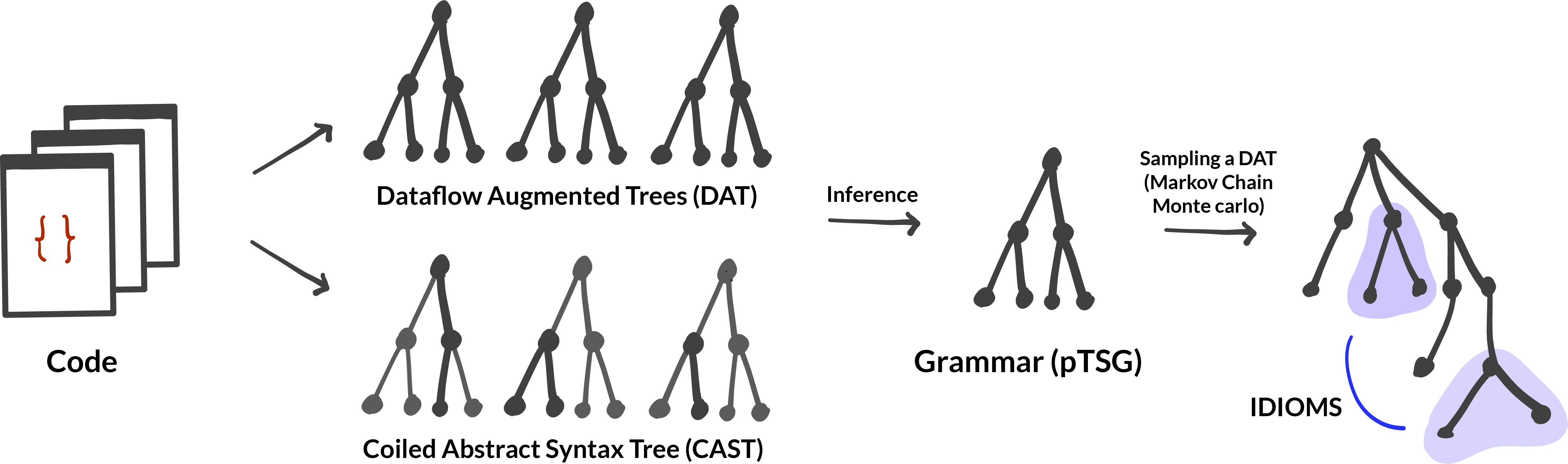}
    \caption{Overview of the steps in mining-based approaches.}
    \label{fig:overview}
\end{figure}

In this work, we propose a new canonicalized dataflow tree representation that overcomes the limitations of prior work listed in Section~\ref{sec:prior_idiom}. 
The upper path of Figure~\ref{fig:overview} provides an overview of \emph{Jezero}, the tool implementing this approach; as is clear, the difference with ~\cite{allamanis2018mining} is that we eschew coiling, and instead work with dataflow augmented trees.
Section~\ref{sec:dataflowtrees} describes the construction of dataflow augmented trees, which is our
new technical contribution, and Section~\ref{sec:stats} gives an overview of the unsupervised idiom learning and sampling approach,
which is the same as in previous
work~\cite{allamanis2014mining}. Note that~\cite{allamanis2014mining} goes directly from code ASTs to pTSG, without any tree augmentation.

The process of using \emph{Jezero} involves the following steps: (1) point
\emph{Jezero} at a corpus that is likely to contain instances of the idiomatic pattern we want to uncover; (2) let \emph{Jezero} mine the patterns and come up with a most suitable one(s) using its ranking heuristics; and (3) use \emph{Jezero} further to point out locations in the code where similar refactoring can be carried out.
%
%

\subsection{Dataflow Augmented Trees}
\label{sec:dataflowtrees}
%
The key insight of \emph{Jezero} is that high-level concepts can be identified as dataflow patterns. Furthermore, these patterns can be captured and represented as canonical trees using an inexpensive dataflow analysis procedure. The problem with representing dataflow information in detail is that there is not enough commonality 
across specific dataflow graphs for a useful semantic pattern to emerge using 
a statistical process of mining patterns.  
This challenge is often referred to as the \emph{sparsity} 
problem~\cite{allamanis2014mining}. 

\emph{Jezero} combats the sparsity problem 
with an abstraction that relies on dataflow information structured in a canonicalized way to capture 
high-level semantic concepts only. Construction of a dataflow augmented tree entails the following steps. First, we extract approximate type information from the AST. Next, we propose a lightweight algorithm that uses the extracted types and information present in the AST to derive flow information as dataflow tables. To mine useful idiomatic patterns, we propose a new tree representation that is amenable to the unsupervised pattern mining algorithm. This canonicalized tree is constructed using information from the dataflow tables. Additionally, we make the following assumptions: (1) mining trees at method level is sufficient to capture refactoring idioms, (2) it is sufficient to keep the control flow structure and collapse
control-free sequence code to a region (an approximation that is often used 
in static analyses~\cite{rosen1988global}), and (3) the side effects, if any, of 
function calls are inconsequential to the dataflow information that we intend to 
capture.


%
\paragraph{\textbf{Static Extraction of Type Information.}}
%
To encode semantic
information, we need to capture the data \textit{type} of each variable. 
However, precise type
information of the variables is often not necessary and can be counterproductive. 
For example, the type of variable \texttt{output} in
Figure~\ref{fig:motivation_eg_1} is \texttt{string} whereas the type of variable 
\texttt{call\_stack\_nodes} in Figure~\ref{fig:motivation_eg_2} is \texttt{vector}. 
Therefore, having precise type information would lead to different patterns. 
Whereas, if the role of those variables in  both cases is to act as a collection, 
we want to only ascribe a \texttt{collection} type to both.

We overcome the need for an expensive analysis algorithm by proposing an approximate
type analysis based on information available in the ASTs. In our
approach, variables are assigned as either \texttt{collection},
\texttt{object} or \texttt{primitive} type.  Each variable is assigned
\texttt{primitive} as the default type and, based on hints from the syntax 
tree, the type may be modified to \texttt{collection} 
or \texttt{object}. For example, if a variable contains a subscript operator, 
it is assigned a \texttt{collection} type. Similarly, if a variable contains 
the arrow operator (or equivalent operators in other programming languages) 
it is assigned an \texttt{object} type. At the end of this procedure, we
have a type table that assigns types for each variable in the method.

As an example, Table~\ref{tab:type_motivation_example_4} summarizes the 
types for the code example in Figure~\ref{fig:motivation_eg_2}. 
Note that these types are particularly useful for identifying map/filter
APIs, and can be tweaked when looking for other API-related patterns.
For instance, \emph{string} types would be necessary when
searching for patterns using string APIs.
\begin{table}[h]
\footnotesize
\centering
\begin{tabular}{@{}ll@{}}
\toprule
\multicolumn{1}{c}{\textbf{Variable}} & \multicolumn{1}{c}{\textbf{Type}} \\ \midrule
call\_stack\_nodes                    & collection                        \\
identifier\_to\_id                    & collection                        \\
identifier                            & primitive                         \\
id                                    & primitive                         \\
nodes                                 & collection                        \\
identifier\_to\_count                 & collection                        \\
this                                  & object                            \\
total\_count                          & primitive                         \\ \bottomrule
\end{tabular}
\caption{Inferred types for the variables in Figure~\ref{fig:motivation_eg_2}.}
\label{tab:type_motivation_example_4}
\end{table}
\paragraph{\textbf{Static Extraction of Dataflow Information.}}
We use dataflow tables to capture data writes and data reads that happen in a code block. To construct these tables, we propose a lightweight analysis that derives dataflow information based on the AST and the type table collected in the previous step.   We mention at the very outset that this dataflow representation is not intended to be \textit{sound}, as needed in 
compiler optimizations.  This choice lets us get away with specific choices that are effective for the purpose at hand. The analysis computes dataflow table $\sigma$ at each control-flow point like \texttt{if}, \texttt{foreach}, etc. $\sigma$ 
encodes the data reads that a variable depends on. Formally, $\sigma \in \mathit{Ref} \rightarrow 2^{\mathit{Ref}}$, where $\mathit{Ref}$ is a tuple containing a canonicalized identifier (\textit{id}) and a variable being referenced.

Identifier for a data write is generated using the variable type and the order of appearance of the write in the current control-flow block. In case of Figure~\ref{fig:motivation_eg_2}, the unique 
identifier for variable $call\_stack\_nodes$ would be $collection\_write\_0$ since it is the first collection variable being written to, although it is the second collection variable in the order of appearance.
$\mathcal{R}$ represents a set of \emph{read references}. Identifier for a data read is
generated using the variable type and the order of appearance in the control-flow block. For the example, in Figure~\ref{fig:motivation_eg_2}, read reference for the variable $id$ would be $(primitive\_1, id)$. 
Examples of flow operations, $f\llbracket \cdot \rrbracket$, include:
\begin{gather*}
    f\llbracket x := y\:op\:z\rrbracket = [\mathit{Ref}_x \rightarrow \{\mathcal{R}_y \cup \mathcal{R}_z\}] \sigma \\
    f\llbracket x := fc(y, z, k) \rrbracket = [\mathit{Ref}_x \rightarrow \{\mathcal{R}_y \cup \mathcal{R}_z \cup \mathcal{R}_k\}] \sigma \\
    f\llbracket \mathit{foreach} (x\:as\:y \Longrightarrow z) \{\}\rrbracket = [\mathit{Ref}_y \rightarrow \mathcal{R}_x; \mathit{Ref}_z \rightarrow \mathcal{R}_x] \sigma
\end{gather*}
The first flow function  calculates $\sigma$ when there is an assignment of an 
expression to a variable. We compute a unique reference $\mathit{Ref}_x$ and compute 
\emph{read references} for each variable in the expression. We then take
a union of these references to update the dataflow table; $\mathit{Ref}_x$ now maps 
to these set of read references. The second flow function is for assignment of the 
return values of a function call to a variable. We compute an unique identifier and 
read references similar to the previous flow function. Note that we assume that for the 
purposes of idiom mining, we can ignore side-effects from function calls. 
The third flow function is for a \texttt{foreach} statement with an empty body. Since we have 
data writes to two variables ($y$ and $z$) we create two unique references $\mathit{Ref}_y$ 
and $\mathit{Ref}_z$. We further compute the read references for each of these variables and update 
$\sigma$. For a given data write, we identify all read dependencies using a fix-point computation. Table~\ref{tab:fixpointdf} illustrates the state of $\sigma$ in two iterations of the 
fix-point computation in Figure~\ref{fig:motivation_eg_2}.

\begin{table}[ht]
    \footnotesize
    \begin{subtable}[h]{.5\textwidth}
      \centering
      \begin{tabular}{p{3cm}|p{5cm}} 
\toprule
\multicolumn{1}{c}{\textbf{Data Write}} & \multicolumn{1}{c}{\textbf{Data Read}}                                                                                     \\ \midrule
(primitive\_write\_0, identifier)                            & (collection\_0, identifier\_to\_id)                                                                                                     \\ \midrule
(primitive\_write\_1, id)                                    & (collection\_0,identifier\_to\_id)                                                                                                       \\ \midrule
(collection\_write\_0, call\_stack\_nodes)                   &  (primitive\_0, identifier), (primitive\_1, id),  (collection\_2,nodes),  (collection\_3, identifier\_to\_count), (object\_0, this), (primitive\_2, total\_count)
\\ \bottomrule
    \end{tabular}
\caption{First Iteration}
    \end{subtable}
    \begin{subtable}[h]{.5\textwidth}
      \centering
      \begin{tabular}{p{3cm}|p{4.5cm}} 
\toprule
\multicolumn{1}{c}{\textbf{Data Write}} & \multicolumn{1}{c}{\textbf{Data Read}}                                                                                     \\ \midrule
(primitive\_write\_0, identifier)                            & (collection\_0, identifier\_to\_id)                                                                                                     \\ \midrule
(primitive\_write\_1, id)                                    & (collection\_0,identifier\_to\_id)                                                                                                       \\ \midrule
(collection\_write\_0, call\_stack\_nodes)                   &  
\textbf{(collection\_0, identifier\_to\_id)}  
(primitive\_0, identifier), (primitive\_1, id),  (collection\_2,nodes),  (collection\_3, identifier\_to\_count), (object\_0, this), (primitive\_2, total\_count)
\\ \bottomrule
\end{tabular}
\caption{Second Iteration.}
    \end{subtable}
    \caption{Figure~\ref{fig:motivation_eg_2}'s Intermediate Dataflow tables ($\sigma$).}
    \label{tab:fixpointdf}
  \end{table}
 
A key aspect of this representation of dataflow tables --- essential to overcome 
the limitations of the approaches detailed in Section~\ref{sec:prior_idiom} --- is
the fact that we propose a new canonicalized label (\textit{id}) for each
dataflow operation. Each label is obtained by concatenating the data type of 
the variable with a number. This canonicalized label helps to overcome the
lexical ordering issues of previous approaches. In particular, we propose 
that each type of data write has its own numbering. For example, primitive writes have their 
own numbering which is incremented whenever there is a data write to a primitive variable. 
This canonicalization allows for interleaved data writes to different types
of variables. While the data writes have special numbering, the data
read references are computed based on their order of appearance. Hence,
the same variable can have a different data read and write reference.
Nested control-flow blocks require construction of $\sigma$ to take into 
account the direction of information flow. There are two choices regarding the 
information flow (1) top-down, where $\sigma$ is carried from the outer to the 
inner code block (2) bottom-up, where $\sigma$ is carried from the inner to the outer code block. 
The choice of information flow influences the type of idioms that we can mine. 
%
%
%
The top-down information flow allows to capture idioms that require context information.
For instance, consider the following code snippet where in the inner loop there 
is a collection variable \texttt{results} populated with the result of function
calls on the items in the \texttt{values} collection. 
\begin{lstlisting}[caption={Nested control-flow block example.},label={lst:nested},captionpos=b,language=PHP]
foreach ($identifiers as $key => $values) {   
    $exp = self::computeExp($key);
    $results = self::computeFirstSecond($key);
    foreach ($values as $item) {
      $results[]= self::computeResult($item, $exp); 
    } 
}
\end{lstlisting}
We cannot directly assign the result of
\texttt{Vec\textbackslash{}map\_with\_key} to \texttt{results}, rather we have to 
\texttt{Vec\textbackslash{}concat} with the items already in \texttt{results}. 
While the top-down information flow can lead to richer idioms, it suffers from two problems: (1) tree sparsity, since no two code blocks share the almost similar context information (2) the learning algorithm proposed by ~\cite{allamanis2018mining} learns context free grammars; however, if we want to use context information, a mining approach that can learn context \textit{sensitive} grammars is needed. 

On the other hand, bottom-up information flow allows us to identify local patterns 
which helps avoid the sparsity problem and is amenable to the grammar learning 
algorithm. Therefore in \emph{Jezero}'s $\sigma$ construction, information flow is 
bottom-up, i.e., from inner to outer code block. In this approach, for each of the 
control flow node, we recursively compute dataflow table for inner block and update 
the outer block using $\psi$ - a merge operator. The operator takes as input two 
dataflow tables and returns a merged dataflow table, formally, $\psi : \sigma_o 
\rightarrow \sigma_i \rightarrow \sigma_m$, where $\sigma_o$ and $\sigma_i$ 
represent the outer and inner-block respectively; and $\sigma_m$ is the merged 
table. We have the following update rules for $\psi$:
\begin{eqnarray*}
    (x, r) \in \sigma_o \land (x, \_) \notin \sigma_i \Longrightarrow
    [x \rightarrow r] \sigma_m \\
    (x, \_) \notin \sigma_o \land (x, \_) \in \sigma_i \Longrightarrow 
    \sigma_m \\
    (x,r1) \in \sigma_o \land (x, r2) \in \sigma_i \Longrightarrow
    [x \rightarrow \\ \mathit{fix} (r1 \cup \{(id_o, v) | (id, v) \in r2 \land v \in \sigma_o \})] \sigma_m 
\end{eqnarray*}
Table~\ref{tab:mergeexample} illustrates the working of the merge operator for the example in Listing~\ref{lst:nested}. \emph{Jezero} starts 
by computing the information for each inner control flow block. Table~\ref{tab:innerregion} illustrates the the dataflow table ($\sigma_i$) for the inner \texttt{foreach} loop. Table~\ref{tab:mergedregion} computes the partial dataflow table ($\sigma_o$) for the outer block without the nested loop. Now we carry out merge using the rules of $\psi$ operator. We retain the first three entries of $\sigma_o$ as per the first rule of $\psi$. Next, we discard the first row of $\sigma_i$ as per the second rule of $\psi$. For write to $results$ variable, we need to merge and update the dataflow from the inner block. We use the third rule of $\psi$ to collect all read references whose variables are declared in the outer context and update the identifier to reflect their position in the outer code block (see Table~\ref{tab:mergedregion}). 

Despite the fact that this bottom-up-only dataflow computation is incomplete, and that traditional context-sensitive analysis may provide sound semantic information, we make this choice purposely as it works well for finding idioms.

\begin{table*}[!htb]
\centering
\resizebox{\textwidth}{!}{%
    \begin{subtable}[h]{.5\textwidth}
      \centering
      \begin{tabular}{p{3cm}|p{5cm}} 
\toprule
\multicolumn{1}{c}{\textbf{Data Write}} & \multicolumn{1}{c}{\textbf{Data Read}}                                                                                     \\ \midrule
(primitive\_write\_0, key)                            
& (collection\_0, identifiers)                                                                                                     \\ \midrule
(collection\_write\_0, values)                                    & (collection\_0,identifiers)                                                                                                       \\ \midrule
(primitive\_write\_1, exp)                                    & (primitive\_0,key)                                                                          \\ \midrule
(collection\_write\_1, results)                   &  (primitive\_0, key)
\\ \bottomrule
\end{tabular}
\label{tab:outerregion}
\caption{Dataflow table of outer region before merge ($\sigma_o$)}
    \end{subtable}%
    \begin{subtable}[h]{.5\textwidth}
      \centering
      \begin{tabular}{p{3cm}|p{4.5cm}} 
\toprule
\multicolumn{1}{c}{\textbf{Data Write}} & \multicolumn{1}{c}{\textbf{Data Read}}                                                                                     \\ \midrule
(primitive\_write\_0, item)                            & (collection\_0, values)                                                                                                     \\ \midrule
(collection\_write\_0, results)                   &  
(collection\_0, values)
(primitive\_0, item), (primitive\_1, exp)
\\ \bottomrule
\end{tabular}
\caption{Dataflow table of inner (\texttt{foreach}) region ($\sigma_i)$}
\label{tab:innerregion}
    \end{subtable}

    \begin{subtable}[h]{.5\textwidth}
      \centering
      \begin{tabular}{p{3cm}|p{5cm}} 
\toprule
\multicolumn{1}{c}{\textbf{Data Write}} & \multicolumn{1}{c}{\textbf{Data Read}}                                                                                     \\ \midrule
(primitive\_write\_0, key)                            
& (collection\_0, identifiers)                                                                                                     \\ \midrule
(collection\_write\_0, values)                                    & (collection\_0,identifiers)                                                                                                       \\ \midrule
(primitive\_write\_1, exp)                                    & (primitive\_0,key)                                                                          \\ \midrule
(collection\_write\_1, results)                   &  (primitive\_0, key)
\textbf{(primitive\_1, exp)}, \textbf{(collection\_1, values)}
\\ \bottomrule
\end{tabular}
\caption{Dataflow table of outer region after merge ($\sigma_m$)}
\label{tab:mergedregion}
        \end{subtable}
}
    \caption{Dataflow tables of outer ($\sigma_o$), inner ($\sigma_i$) and merged ($\sigma_m$) regions for the code example in Listing~\ref{lst:nested} }
    \label{tab:mergeexample}
  \end{table*}
  

%
\paragraph{\textbf{Tree Representation.}}
The pattern mining algorithm we use (see Section~\ref{sec:stats}) learns tree fragments given a context free grammar. Therefore, having as a starting point the dataflow information captured as tables, we 
need a suitable tree representation that is amenable to (tree) pattern mining. In this work, we replace a control-free sequence of statements with trees that capture dataflow information. This tree contains information about the data writes and data reads that happen in a code block. To ensure that these trees are compatible with the underlying learning technique, we propose the following canonicalized tree representation.

\begin{figure*}[ht]
    \centering
    \includegraphics[width=0.90\textwidth]{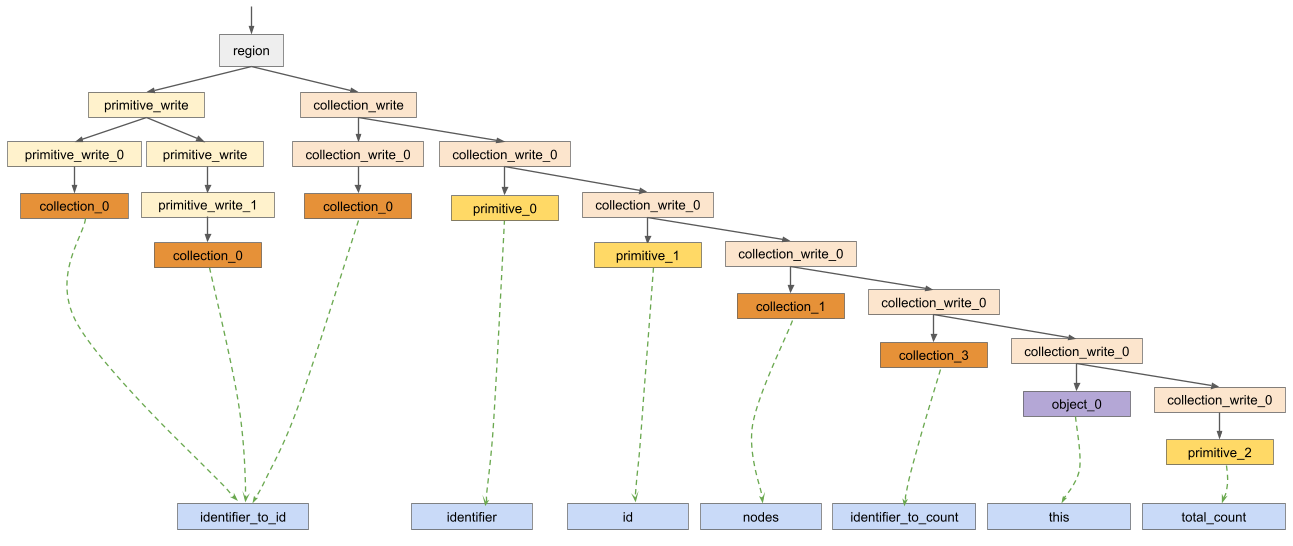}
    \caption{Region tree representing the dataflow operations for the code example in Figure~\ref{fig:motivation_eg_2}. Due to space limitations, we show a compact version of the AST instead; e.g., we collapse ids and rules into one node with the rule suffixed with id as label.}
    \label{fig:motivation_eg_4_region}
\end{figure*}

To be capable of 
differentiating between distinct data writes, our proposed tree representation always contains a set of (distinct) 
child nodes that represent data writes to \texttt{collection}, \texttt{primitive}, or \texttt{object} type. This design choice helps in overcoming the 
limitation with different number of write statements, since the 
learning algorithm will learn to retain only the common data write 
pattern (i.e., the common subtree).  
Additionally, to account for the dataflow in the loop header, we add 
primitive write dataflow nodes which 
shows the flow from the collection being iterated over to the loop 
header variables. \emph{Jezero} models dataflow tables as trees using the following grammar:
\begin{bnf*}
\bnfprod{region}{
  \begin{bnfsplit}
    \bnfpn{$primitive\_write$} 
    \\ \bnfts{       } \bnfpn{$collection\_write$} \bnfpn{$object\_write$}
  \end{bnfsplit}
}\\
\bnfprod{$primitive\_write$}{
  \begin{bnfsplit}
    (\bnfpn{$write\_region$} \bnfpn{$primitive\_write$})*
  \end{bnfsplit}
}\\
\bnfprod{$collection\_write$}{
  \begin{bnfsplit}
    (\bnfpn{$write\_region$} \bnfpn{$collection\_write$})*
  \end{bnfsplit}
}\\
\bnfprod{$object\_write$}{
  \begin{bnfsplit}
    (\bnfpn{$write\_region$} \bnfpn{$object\_write$})*
  \end{bnfsplit}
}\\
\bnfprod{$write\_region$}{
  \begin{bnfsplit}
    (\bnfpn{id} \bnfpn{$read\_region$} \bnfpn{$write\_region$})*
  \end{bnfsplit}
}\\
\bnfprod{$read\_region$}
{\bnfpn{id} \bnfpn{$variable\_name$}}\\
\end{bnf*}
Figure~\ref{fig:motivation_eg_4_region} illustrates the region node for the code example in 
Figure~\ref{fig:motivation_eg_2}. For this particular example, we have data writes to two 
primitive variables in the loop header from the collection variable 
\texttt{identifier\_to\_id} which is captured in the 
\texttt{primitive\_write} subtree. In the body of the loop, there is 
a write to a collection variable \texttt{call\_stack\_nodes}. This 
dataflow is represented as a child node in the 
\texttt{collection\_write} subtree. Since this is the first collection 
write in the loop body it is referenced with 
\texttt{collection\_write\_0} and the data reads from 
\texttt{identifier}, \texttt{id}, \texttt{nodes}, and
\texttt{identifier\_to\_count} are represented as a right balanced subtree. The fix-point operation identifies that the variable write to \texttt{call\_stack\_nodes} also depends on data read from \texttt{identifier\_to\_id} since \texttt{identifier} and \texttt{id} data write depends on it.
The canonicalization can be seen in the data read reference for \texttt{call\_stack\_nodes}, 
which is \texttt{collection\_1} whereas the data write reference for the same variable is
\texttt{collection\_write\_0}. Construction of similar canonicalized trees for other code snippets in Figure~\ref{fig:motivation_idiom_examples} helps identify the common pattern mentioned in Section~\ref{sec:introduction}. The pattern identified from these trees is that \texttt{foreach} region contains a \texttt{collection\_write} to a \texttt{collection\_0} variable with data read from \texttt{primitive\_0}, \texttt{primitive\_1} and \texttt{collection\_0}. 

The proposed representation captures information flow between variables in addition to the variable mutability, whereas CASTs~\cite{allamanis2018mining} capture only variable mutability in their region nodes. Canonicalized labels for each type of data write maps the first data write in Figure~\ref{fig:motivation_eg_1},~\ref{fig:motivation_eg_2}, ~\ref{fig:motivation_eg_3} to the same variable reference whereas CASTs maps them to different references (see Figure~\ref{fig:semantic_loops_motivation_examples}). \emph{Jezero}'s tree representation arranges the dataflow information such that a \textit{top-down} mining algorithm (Section~\ref{sec:stats}) can efficiently extract frequent subtrees (i.e., frequent flow patterns) from large sets of trees. 

\subsection{Mining Idioms}
\label{sec:stats}
Allamanis and Sutton~\cite{allamanis2014mining} propose probabilistic tree substitution 
grammars to infer and capture code idioms. A tree substitution grammar (TSG) is an extension 
to a context-free grammar (CFG), in which productions expand into tree fragments. Formally, a TSG is a tuple $G = (\Sigma,N,S,R)$, where $\Sigma$ is a set of terminal symbols, $N$ is a set of nonterminal symbols, $S \in N$ is the root of the nonterminal symbol and $R$ is a set of productions. In case of TSG, each production $r \in R$ takes the form $X \to T_X$ , where $T_X$ is a tree fragment rooted at the nonterminal $X$. 
    
    
    

The way to produce a string from a TSG is to begin with a tree containing $S$ only, and recursively
expand the tree top-to-bottom, left-to-right as in CFGs --- the difference is that some rules can
increase the height of the tree by more than 1.
A pTSG augments a TSG with probabilities, in an 
analogous way to a probabilistic CFG (pCFG). Each 
tree fragment in the pTSG can be thought of as 
describing a set of context-free rules that are used 
in a sequence. Formally, a pTSG is $G = 
(\Sigma,N,S,R,\sigma),$ which augments a TSG with 
$\sigma$, a set of distributions $P_{TSG}(T_X|X)$, 
for all $X \in N$, each of which
is a distribution over the set of all rules $X \to T_X$ in $R$ that have left-hand side $X$.

The goal of our mining problem is to infer a pTSG in which every tree fragment represents a code 
idiom. Given a set of trees ($T_1, ..., T_n$) for pTSG learning, the key factor that determines 
model complexity is the number of fragment rules associated with each nonterminal. If the model 
assigns too few fragments to a non-terminal, it will not be able to identify useful patterns
(\textit{underfitting}); on the other hand, if it assigns too many fragments, then it can simply 
memorize the corpus (\textit{overfitting})~\cite{allamanis2018mining}. Furthermore, we do 
not know in advance how many fragments are associated with each non-terminal. Non-parametric Bayesian 
statistics~\cite{murphy2012machine,gelman2013bayesian} 
provide a simple, yet powerful, method to manage this trade-off for cases where the number of 
parameters is unknown. In this work, we use the nonparametric Bayesian inference methods proposed by Allamanis and Sutton~\cite{allamanis2014mining} to mine refactoring idioms. 
To infer a pTSG $G$ using Bayesian inference, we first compute 
a probability distribution over probabilistic grammars, $P(G)$. This distribution is bootstrapped by estimating the maximum likelihood from our training corpus.  While this gives distribution 
over full trees, we require the distribution over fragments. This is defined as 
$P_0(T) = \prod_{r \in T} P(r)$, where $r$ ranges over the set
of productions that are used within $T$. The specific prior distribution that we use is Dirichlet process. 
The Dirichlet process is specified by a \textit{base measure}, which is the fragment
distribution $P_0$, and a \textit{concentration parameter} $\alpha \in \mathbb{R}^+$ 
that controls the rich-get-richer effect.
Given $P_0$ and prior distribution, we apply Bayes' rule to obtain posterior 
distribution. The posterior Dirichlet process pTSG is characterized by a finite set of tree fragments for 
each non-terminal. To compute this distribution, we resort to approximate inference based on Markov 
Chain Monte Carlo (MCMC)~\cite{liang2010type}. Specifically, we use Gibbs sampling to sample the posterior distribution over 
grammars.

At each sampling iteration, \emph{Jezero} samples the trees from the corpus and for each 
node in the tree it \textit{decides} if it is a root or not based on posterior probability. \emph{Jezero} adds trees to the sampling corpus and adds tree fragments to the sample grammar based on whether the fragments are root (denoted by $z_t = 1$) or not. Next, for each tree node $T_t$, \emph{Jezero} identifies the parent $T_s$ whose $z_s = 1$. Based on the current node and its root parent, \emph{Jezero} samples it to decide whether to merge them as a single fragment or to separate them into different fragments. To do this, \emph{Jezero} computes the probability of the joint tree (node $T_t$ and parent $T_s$), and the split probabilities. Based on a threshold it either splits the node and tree into fragments or merges them into one fragment; 
\[
Pr(z_t = 0) = \frac{Pr_{post}(T_{join})}{Pr_{post}(T_{join}) + Pr_{post}(T_s) \cdot Pr_{post}(T_t)}
\]

\noindent
\[
Pr_{post}(T) = \frac{count(T) + \alpha \cdot P_0(T)}{count(h(T)) + \alpha}
\]

\noindent
$T_{join} = T_t \cup T_s$, $h$ is the root of the fragment, and count is the number of 
times that a tree occurs as a fragment in the corpus, as determined by the current values of
$z_t$. Once the sampling is complete, \emph{Jezero} orders the grammar based on the production probability and filter out those rules that have probability is less than $0.5$.

    
    



    

In MCMC it is essential that there is good mixing of samples, hence \emph{Jezero} visits the  
the trees in the corpus and their nodes in different orders to further introduce randomness. We seed the sampling 
process by annotating randomly 90\% of the nodes with $z_t = 1$\footnote{Other annotation
values (namely, 40\% and 60\%) yield similar results at the cost of 3x slowdown in execution time.}.
Furthermore, it incrementally adds trees to the corpus to compute the grammar. \emph{Jezero} repeats this process for $50$ iterations to identify the posterior distribution over fragments, which are then 
returned as idioms. We further experimented with $100$ iterations but no changes in the evaluation
scores for the APIs were observed.
\begin{figure}[t!]
  \centering
          \includegraphics[width=0.35\textwidth]{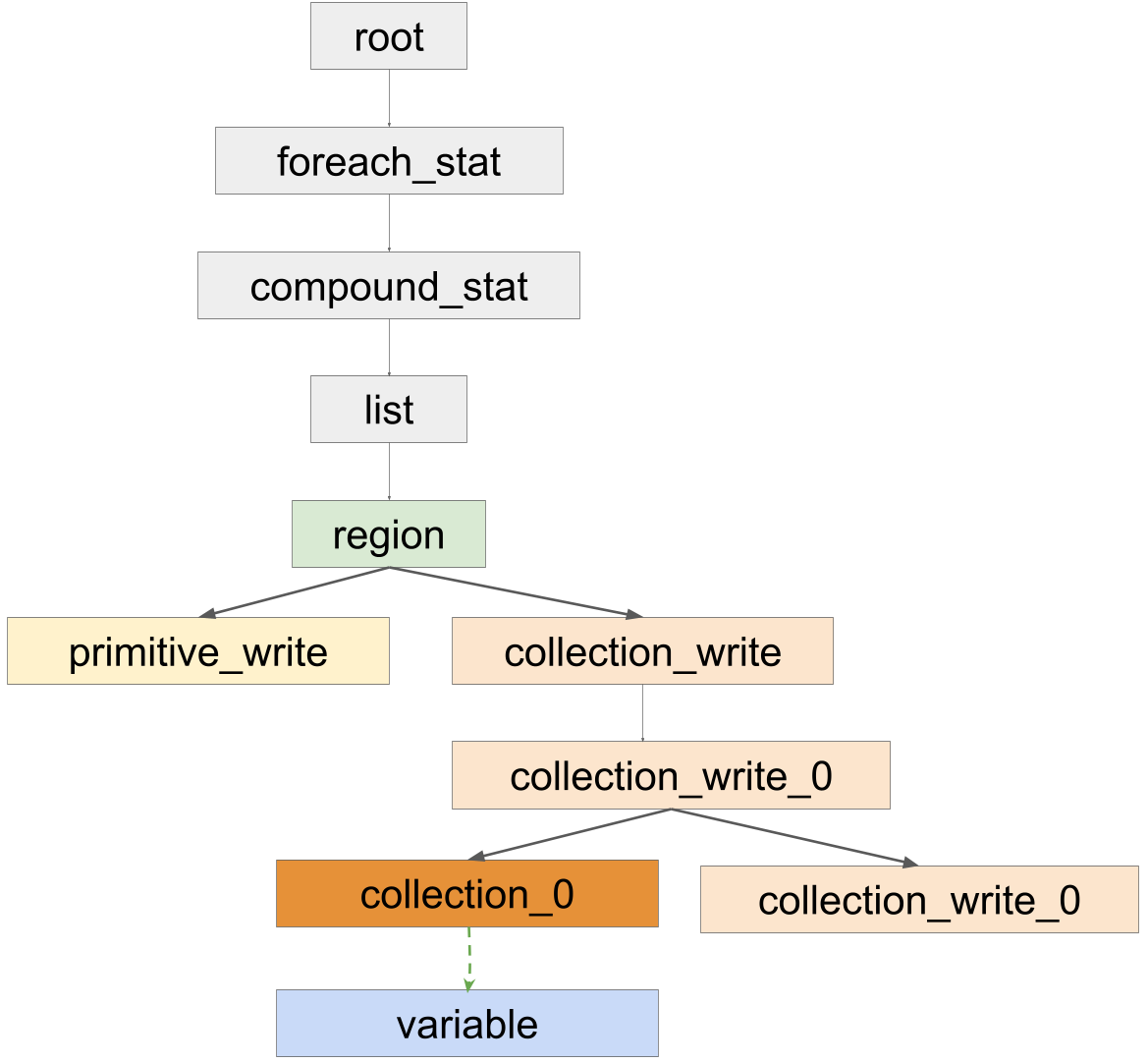}
          \caption{Example idiom mined by \emph{Jezero}: Top-1 idiom for API Vec\textbackslash{}map\_map\_with\_key. An example of a code snippet for this idiom is in Figure~\ref{fig:motivation_eg_1}.}
          \label{fig:top1vecmapwithkey}
%
%
%
      \label{fig:comparingidioms}
\end{figure}

Figure~\ref{fig:comparingidioms} shows the top-1 idiom mined by \emph{Jezero} for the 
Vec\textbackslash{}map\_map\_with\_key API. 
\textbf{Note how this is prefix of the tree shown in Figure~\ref{fig:motivation_eg_4_region}: the key advantage of the canonicalized representation.} 
Despite its shallow nature, in Section~\ref{sec:evaluation}, we show that
this idiom is very effective in identifying refactoring opportunities. In particular, 
according to this idiom, the most common dataflow pattern is a loop with a write to 
the first collection variable, and it depends on the loop iteration variable.
The reason for \emph{Jezero} to return shallow idioms is attributed to the Gibbs sampling process. 
Finding deeper idioms is possible by tuning the Gibbs sampling hyper-parameters (it remains,
however, for future work).
\subsection{Idiom Ranking}
\label{sec:ranking}
The mining process may compute a large number of idioms, and therefore, we need 
a mechanism to rank the idioms based on their usefulness in identifying 
refactoring patterns. We identify three different ranking schemes that
could help surfacing interesting patterns. Before ranking the idioms, we first 
prune away trivial idioms based on two rules: (1) remove idioms that have been 
seen less than a minimum number of times (\texttt{c\_min}); and (2) remove idioms 
that have less than a minimum number of nodes 
(\texttt{n\_min})~\cite{allamanis2014mining}. 

\paragraph{\textbf{Ranking based on Coverage (C)}}
Idiom coverage~\cite{allamanis2014mining} is computed as the percent of source 
code dataflow trees that are matched to the mined idiom. Coverage ranges 
between $0$ and $1$, indicating the extent to which the mined idioms exist in a 
piece of code.
\paragraph{\textbf{Ranking based on Information-theoretic Measure (CE)}}
To maximize information content and coverage, Allamanis \textit{et 
al.}~\cite{allamanis2018mining} score an idiom by multiplying coverage and
cross-entropy gain. Cross entropy
gain measures the expressivity of an idiom and averages log-ratio of the posterior
pTSG probability of the idiom over the probability yielded by the basic pCFG. 
This ratio measures how much each idiom ``improves'' upon the base pCFG distribution.
\paragraph{\textbf{Ranking based on Jaccard Similarity (IOU)}}
In addition to the two ranking schemes from prior work, we propose a scheme
based on the average \textit{area} an idiom covers when matching a code location. 
The score $s$ for an idiom $i$ is computed as follows:
\begin{equation}
    s(i) = weight(i) \times support(i) \times IoU(i)
\end{equation}

\noindent
where $weight$ is a heuristic function that returns a value between $0$ and $1$ based 
on the pre-assigned weight of the root of the idiom  (e.g. \texttt{foreach}, \texttt{if} have higher weight than \texttt{primitive\_write}). Function $support$ computes 
the number of trees whose subtree is the idiom $i$. The term $IoU(i) \in [0,1]$ 
(\textit{intersection over union}) measures the average number of 
nodes that overlap an idiom and its supports. 
\section{Evaluation}
\label{sec:evaluation}
This section details the empirical evaluation of \emph{Jezero} on the task 
of learning refactoring idioms for a diverse set of APIs from the Facebook Hack codebase. Further, for each API, we measure the effectiveness of the mined idiom in identifying known and potential refactoring locations.

\begin{figure}[ht]
    \centering
    \includegraphics[scale=0.20]{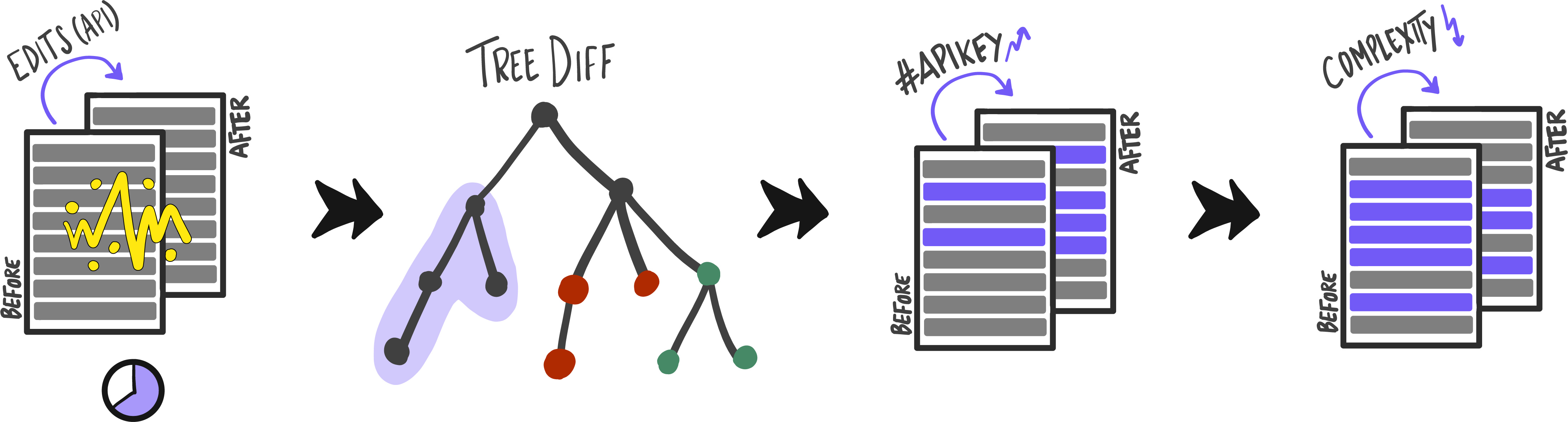}
    \caption{Phases of the dataset construction.}
    \label{fig:exp_setup}
\end{figure}

\paragraph{\textbf{Dataset for Mining}} As distinct APIs have different refactoring
patterns, per API, we construct a dataset of patterns using historical change data. We call
these changes \emph{edits} and they contain a \textit{before} and \textit{after} version of
changed source file(s). We first scrape edits in a given time 
interval and API from Facebook's code repository and construct a dataset for mining 
refactoring patterns using \textit{before} versions of the edits. 
However, this suffers from a drawback that edits collected using this 
approach would, most likely, contain excessive noise, i.e., changes that are not relevant to the
refactoring changes. Hence, we opted for a relatively inexpensive way to prune out irrelevant 
edits. We propose the following three-fold heuristic filter, as shown in 
Figure~\ref{fig:exp_setup}: (1) first we compute a ``treediff'' 
of each edit using the GumTree algorithm \cite{Falleri2014FineGrained} and remove method 
trees that were not modified or did not see an introduction of the API we are investigating,
(2) we then collect code edits whose API keyword occurrence in after version is higher
compared to before, (3) we further filter edits where the cyclomatic complexity of
\textit{before} is greater than the \textit{after}. These heuristics are not meant to 
end up with the actual refactorings exclusively, but to increase the chances of each \textit{before}-\textit{after} pair being a valid refactoring. We believe that the
resulting diversity in the dataset helps prevent overfitting. 
On average, for the APIs in Table~\ref{tab:eval_f1_score} the initial dataset contains 
$21,147$ trees rooted at the method level (i.e., number of methods --- average per API --- 
found before the three-fold heuristic). After the pruning stage, on average, the dataset 
contains $1,347$ training trees rooted at method level ($1,347$ trees for
Vec\textbackslash{}map\_with\_key; $1,151$ 
trees for Vec\textbackslash{}gen\_filter; $2,198$ trees for Dict\textbackslash{}filter; $693$ 
trees for Dict\textbackslash{}map\_with\_key).

\paragraph{\textbf{Experimental Setup}}

We evaluate the effectiveness of \emph{Jezero} in two settings. 
First, we measure the  accuracy of the proposed approach on a manually constructed validation
set, containing true refactoring opportunities and non-opportunities.

Second, 
we measure the performance of \emph{Jezero} in identifying refactoring opportunities 
in the entire codebase. For each API, we sample idioms for $50$ iterations and with 
a concentration parameter value of $5.0$. Furthermore, we have pruned rare 
(\texttt{c\_min} $ = 2$) and small (\texttt{n\_min} $ = 6$) idioms. The observations 
reported are the result of running \emph{Jezero} for 88 hours on an Intel Core 
Processor i7-6700 CPU @2.39GHz with 57GB RAM. Note that this is the one-time cost to 
train the 4 different APIs. Moreover, for each API, the training takes about the same amount (~22hours). Prediction time is just a few milliseconds to identify matching locations.



\paragraph{\textbf{Effectiveness in identifying known refactoring.}}
To measure the accuracy of the proposed approach, for each API, we manually construct a
ground truth dataset (1) using manually confirmed refactoring locations in the historical change 
data and (2) manually identified potential refactoring locations from a set of files
sampled from the current version of the codebase. These locations from the current 
codebase are included to get a wider variety of code samples.

The constructed evaluation dataset contains $431$ trees\footnote{This is the average number of trees randomly sampled from the $1,347$ mentioned before, as well as files 
sampled from the current codebase}, on average, where $27$ 
trees out of these are true refactoring locations\footnote{27 is the number (average 
per API) of trees that were manually verified to be true refactoring locations.} (see Figure~\ref{tab:eval_f1_score}). 
Note that not all $431$ trees are true refactorings. This happens because, other than
the true refactoring method, an \emph{edits}'s \textit{before} version in our dataset may 
contain several methods with loopy code that is similar to the idiom. Hence, a manual check 
of the trees revealed that $27$ were actual refactorings.

We compare \emph{Jezero} with \emph{Haggis}, the AST-based idiom mining proposed by Allamanis et.al~\cite{allamanis2014mining}. \emph{Haggis} does not have the dataflow augmentation that \emph{Jezero} has, but is otherwise identical to \emph{Jezero}.  We could not compare with semantic idiom mining based on coiling~\cite{allamanis2018mining} because the latter requires annotations or dynamic analysis.

We measure the effectiveness of the proposed approach in identifying refactoring
locations by comparing precision, recall, and the F1 scores of idioms produced by the 
baseline AST-based approach (Haggis) vs \emph{Jezero}. 
Table~\ref{tab:eval_f1_score} shows the accuracy results for four API refactorings 
when using top-1 idiom
using the intersection over union ranking scheme. On average \emph{Jezero's} F1 score 
is significantly better than the 
baseline on all APIs. This indicates the effectiveness of the proposed static analysis 
based tree representation. Additionally, the differences between the APIs reflect how
well the new tree representation can identify diverse patterns. 

\begin{table}[ht]
\centering
\resizebox{\columnwidth}{!}{%
\begin{tabular}{@{}lrcc@{}}
\toprule
& \multicolumn{1}{l}{\textbf{Eval trees}} & \multicolumn{1}{c}{\textbf{Haggis}} & \multicolumn{1}{c}{\textbf{\emph{Jezero}}}\\ 
& (true/total) & F1/precision/recall & F1/precision/recall\\ \midrule
\textbf{Vec\textbackslash{}map\_with\_key} & 36/437 & 0.14/0.43/0.08 & 0.71/0.74/0.69                                  \\
\textbf{Vec\textbackslash{}gen\_filter}     & 22/535 & 0.0/0.0/0.0 & 0.50/0.39/0.72                                  \\
\textbf{Dict\textbackslash{}filter}         & 12/324 & 0.0/0.0/0.0 & 0.56/0.54/0.58                                  \\
\textbf{Dict\textbackslash{}map\_with\_key} & 39/426 & 0.19/0.45/0.13 & 0.51/0.53/0.49                                  \\ \bottomrule
\textbf{Average} & 27 / 431 & 0.08/0.22/0.05 & 0.57/0.55/0.62 \\
\bottomrule
\end{tabular}
}
\caption{Performance of \emph{Jezero} vs. Haggis.}
\label{tab:eval_f1_score}
\end{table}
\paragraph{\textbf{Effectiveness in identifying refactoring opportunities.}}
In this experiment, we measure the performance of \emph{Jezero} in identifying potential 
refactoring locations on Facebook's codebase with $13770$ Hack methods, spread over $1501$ 
files. For each API, we identify matching locations using the top-1 idiom from \emph{Jezero} 
and \emph{Haggis}. Table~\ref{tab:www_eval} summarizes the number of matching locations for
each API. \emph{Jezero} matches $807$ locations (202 on average; i.e., 
$1.5$\% of the trees rooted at loop headers), 
whereas Haggis only matches $23$ locations in all ($6$ on average); \emph{Haggis} fails to
identify \emph{any} refactoring opportunities in the case of 
\texttt{Vec\textbackslash{}gen\_filter} and \texttt{Dict\textbackslash{}filter}. 

Furthermore, to identify the precision of the matched locations, we manually inspect all locations of \emph{Haggis} and a random subset of ($100$) locations for each API returned by \emph{Jezero}.
\begin{table}[ht!]
\footnotesize
\centering
\begin{tabular}{@{}lrr@{}}
\toprule
                                            & \multicolumn{1}{l}{\textbf{\#Matching Locations}} & \textbf{Precision} \\
                                            & \textbf{\emph{Jezero} / Haggis} &  \textbf{\emph{Jezero} / Haggis} \\ \midrule
\textbf{Vec\textbackslash{}map\_with\_key}  & 260 / 1  & 0.91 / 1.00   \\
\textbf{Vec\textbackslash{}gen\_filter}     & 247 / 0  & 0.41 / 0.00   \\ 
\textbf{Dict\textbackslash{}filter}         & 134 / 0  & 0.39 / 0.00   \\ 
\textbf{Dict\textbackslash{}map\_with\_key} & 166 / 22 & 0.68 / 0.91  \\ 
\bottomrule
\textbf{Average} & 202 / 6 & 0.60 / - \\
\bottomrule
\end{tabular}
\caption{Top-1 idioms' matching locations in the wild.}
\label{tab:www_eval}
\end{table}
Note that since we do not have a dataset of locations that should match in the internal codebase, no measures of recall are reported. On average \emph{Jezero} has a precision of $0.60$, which is an 
encouraging number. 
%
\emph{Haggis}'s average precision is inconsequential because of low absolute numbers. 

\textbf{In summary, \emph{Jezero} not only mined  patterns
in an unsupervised way, those mined patterns were extremely productive in locating opportunities in the ``wild''.}


\begin{table}[ht]
\footnotesize
\centering
\begin{tabular}{@{}lrrr|rrr@{}}
\toprule
                                            & \multicolumn{3}{c|}{\textbf{Haggis}}                                                                     & \multicolumn{3}{c}{\textbf{\emph{Jezero}}}                                                              \\ 
                                            & \multicolumn{1}{c}{\textbf{C}} & \multicolumn{1}{c}{\textbf{CE}} & \multicolumn{1}{c|}{\textbf{IOU}} & \multicolumn{1}{c}{\textbf{C}} & \multicolumn{1}{c}{\textbf{CE}} & \multicolumn{1}{c}{\textbf{IOU}} \\ \midrule
\textbf{Vec\textbackslash{}map\_with\_key}  & 0 & 0.14 & 0.14 & 0.71 & 0.71 & 0.71  \\
\textbf{Vec\textbackslash{}gen\_filter}     & 0 & 0 & 0 & 0.50 & 0.45 & 0.50 \\
\textbf{Dict\textbackslash{}filter}         & 0 & 0 & 0 & 0.55 & 0.55 & 0.55 \\
\textbf{Dict\textbackslash{}map\_with\_key} & 0.19 & 0.19 & 0.19 & 0.50 & 0.50 & 0.51
\\ \bottomrule
\end{tabular}
\caption{F1 performance of different ranking schemes.}
\label{tab:ranking}
\end{table}
\paragraph{\textbf{Effectiveness of different ranking strategies.}}
For each API, we rank the mined idioms using the ranking schemes discussed in Section~\ref{sec:ranking} and
compute the F1 score based on top-1 idiom. Table~\ref{tab:ranking} summarizes the results. We can see that 
F1 scores are almost identical across the different ranking schemes. This observation is consistent with 
both \emph{Haggis} and \emph{Jezero}. Therefore any ranking strategy is well suited for both the approaches.
%

%

%
\section{Threats to Validity}
Regarding internal validity,
the effectiveness of parameters may depend on the extent and nature of 
codebase used. To mitigate this risk we have experimented with a 
combination of parameters and ranking schemes. However, we have not 
systematically explored every combination of parameters in our 
experiments. Hence, other combinations may work better to other 
systems. In terms of external validity, the proposed approach has only 
been evaluated using a codebase developed by a single company (albeit 
a large codebase). Furthermore, the approach has been evaluated on the 
Hack programming language, and may only generalize the results to other 
programming languages with prudence. To mitigate this threat, as future 
work, we will investigate the effectiveness of our approach on other 
languages and codebases.

\section{Limitations and Future Work} 
The dataflow trees we generate are type agnostic. Therefore, different APIs 
could have similar idiomatic patterns. For example, we observe that top-3 
idioms of \texttt{Vec\textbackslash{}gen\_filter}, \texttt{Dict\textbackslash{}filter} 
are identical. Other APIs that are expected to yield identical idioms are 
\texttt{Vec\textbackslash{}map\_with\_key} and  
\texttt{Dict\textbackslash{}map\_with\_key}. To improve the precision of the proposed 
approach we can add type information (1) while generating the tree representation, or 
(2) use it to disambiguate APIs at prediction time. Moreover, the current dataflow trees are rather general --- e.g., no information about if-expressions are captured. Adding
additional information for, e.g., the variable references in the \texttt{if} condition, 
will likely help mine better idioms. 

Additionally, the proposed tree canonicalization was influenced by the 
idiom mining machinery which identifies contiguous patterns from trees. Capturing information about variables outside a local code block makes it a graph mining problem. To overcome this, we can introduce predicates, such as \texttt{contains,before,after}, and construct a tree based on this grammar. However, this might lead to a computationally expensive sampling approach.  

Furthermore, the dataflow trees are generated using a bottom-up approach where the information flow is in one direction, from 
the inner to the outer code block. The design choice we made to capture local patterns is not
suitable, for example, for patterns that depend on context information from the outer 
region. This has been discussed in Section~\ref{sec:dataflowtrees}, and illustrated in 
Listing~\ref{lst:nested}.

Finally, refactoring opportunities identified by \emph{Jezero} may not be good
candidates for actual refactoring, due to the replacement API being less performant
or readable. Therefore, we do not plan to automatically apply refactorings detected
by \emph{Jezero}, and instead surface suggestions to developers during the code review process.

\section{Related Work}
We have discussed the work of Allamanis \textit{et al.}~\cite{allamanis2014mining, allamanis2018mining} on idiom mining through the paper. 
To recap, \emph{Jezero} differs from their work in two ways. First, prior work uses annotations or dynamic analysis to capture semantic properties, while \emph{Jezero} instead uses lightweight static analysis, which is more readily available. Second, \emph{Jezero} constructs a canonicalized tree representation that captures dataflow in a control-flow block which provides richer information that we need for our setting of identifying latent refactorings. 

Code clone detection ~\cite{li2004cp,kamiya2002ccfinder,sajnani2016sourcerercc,deckard,roy2008nicad} techniques are related to idiom mining, as the goal is to identify highly similar code blocks. 
Rattan \textit{et al.}~\cite{rattan2013software} identify several clone detection techniques that use syntax and semantics of a program~\cite{baxter1998clone, koschke2006clone}. Code idiom mining proposed in this work searches for frequent as opposed to maximally identical subtrees as with clone detection techniques. Semantic code search techniques~\cite{premtoon2020semantic, david2014tracelet, stolee2014solving, sivaraman2019active} are also related to idiom mining, since they utilize type~\cite{reiss2009semantics}, data, and control flow~\cite{premtoon2020semantic,wang2010matching} information for identifying clones. Our approach differs in two ways; (1) code search requires the user to provide a search pattern, whereas \emph{Jezero} infers such a pattern (2) search techniques that infer a pattern~\cite{sivaraman2019active} leverages active learning while we use nonparametric Bayesian methods. Another related area is API mining~\cite{zhong2009mapo,wang2013mining,acharya2007mining,galenson2014codehint,mandelin2005jungloid}. However, this problem is significantly different from idiom mining since it tries to mine sequences or graphs~\cite{gu2016deep, mou2016convolutional,nguyen2009graph} of API method calls, usually ignoring most features of the language. 
API protocols can be considered a type of semantic idiom; therefore, idiom mining is a general technique for pattern matching and can be specialized to API mining by devising appropriate tree representations.

In recent years, we have seen an emerging trend of tools and techniques that synthesize
program transformations from examples of code 
edits~\cite{bader2019getafix,meng2013lase,miltner2019fly,polozov2015flashmeta,rolim2017learning,rolim2018learning,gao2020feedback}.
%
%
The synthesized transformation should satisfy the given examples while producing correct edits on unseen inputs. Existing approaches have addressed
this in different ways. Sydit~\cite{meng2011systematic} and
LASE~\cite{meng2013lase}, are only able to generalize variables names,
methods and fields. Moreover, the former only accepts one example and
synthesizes the transformation using the most general generalization, whereas the latter accepts multiple examples and synthesizes the transformation using the most specific
generalization. This approach is also the approach taken by Revisar~\cite{rolim2018learning} 
and Getafix~\cite{bader2019getafix}. ReFazer~\cite{rolim2017learning, gao2020feedback} learns a set of transformations consistent with the examples and uses a set of heuristics to rank the transformations in order of likelihood to be correct. 
%
While these techniques learn transformations from the provided examples, \emph{Jezero's} main focus is on the detection of statistically significant patterns directly from a corpus, and then pointing out likely opportunities for refactoring.  On a different note, many of these tools can also benefit from the dataflow augmented tree structure that we introduced that makes the common semantic pattern manifest.


%
\section{Conclusions}
We propose \emph{Jezero}, a scalable, lightweight technique that is capable of surfacing semantic idioms from large codebases. Under the hood, \emph{Jezero} 
extends the abstract syntax tree with canonicalized dataflow trees and 
leverages a well-suited a nonparametric Bayesian method to mine the semantic idioms.

Our experiments on Facebook's Hack code shows \emph{Jezero}'s clear
advantage, as it was significantly more effective than a baseline that did 
not have the dataflow augmentation in being able to find refactoring
opportunities from unannotated legacy code.
On a randomly drawn 
sample containing $13770$ Hack methods,
\emph{Jezero} found matches at 1.5\% locations
among which, the precision of finding a real refactoring
opportunity was $0.60$.

We expect the ideas in \emph{Jezero} to carry over to other languages such 
as Python, as it provides ways to express idiomatic code.

\balance
\bibliographystyle{plain}
\bibliography{draft}

\begin{thebibliography}{10}

\bibitem{acharya2007mining}
Mithun Acharya, Tao Xie, Jian Pei, and Jun Xu.
\newblock Mining api patterns as partial orders from source code: from usage
  scenarios to specifications.
\newblock In {\em Proceedings of the the 6th joint meeting of the European
  software engineering conference and the ACM SIGSOFT symposium on The
  foundations of software engineering}, pages 25--34, 2007.

\bibitem{allamanis2018mining}
Miltiadis Allamanis, Earl~T Barr, Christian Bird, Premkumar Devanbu, Mark
  Marron, and Charles Sutton.
\newblock Mining semantic loop idioms.
\newblock {\em IEEE Transactions on Software Engineering}, 44(7):651--668,
  2018.

\bibitem{allamanis2014mining}
Miltiadis Allamanis and Charles Sutton.
\newblock Mining idioms from source code.
\newblock In {\em Proceedings of the 22nd ACM SIGSOFT International Symposium
  on Foundations of Software Engineering}, pages 472--483, 2014.

\bibitem{bader2019getafix}
Johannes Bader, Andrew Scott, Michael Pradel, and Satish Chandra.
\newblock Getafix: Learning to fix bugs automatically.
\newblock {\em Proceedings of the ACM on Programming Languages},
  3(OOPSLA):1--27, 2019.

\bibitem{baxter1998clone}
Ira~D Baxter, Andrew Yahin, Leonardo Moura, Marcelo Sant'Anna, and Lorraine
  Bier.
\newblock Clone detection using abstract syntax trees.
\newblock In {\em Proceedings. International Conference on Software Maintenance
  (Cat. No. 98CB36272)}, pages 368--377. IEEE, 1998.

\bibitem{cohn2010inducing}
Trevor Cohn, Phil Blunsom, and Sharon Goldwater.
\newblock Inducing tree-substitution grammars.
\newblock {\em The Journal of Machine Learning Research}, 11:3053--3096, 2010.

\bibitem{david2014tracelet}
Yaniv David and Eran Yahav.
\newblock Tracelet-based code search in executables.
\newblock {\em Acm Sigplan Notices}, 49(6):349--360, 2014.

\bibitem{Falleri2014FineGrained}
Jean-R{\'e}my Falleri, Flor{\'e}al Morandat, Xavier Blanc, Matias Martinez, and
  Martin Monperrus.
\newblock {Fine-grained and Accurate Source Code Differencing}.
\newblock In {\em {Proceedings of the International Conference on Automated
  Software Engineering}}, pages 313--324, V{\"a}steras, Sweden, 2014.
\newblock update for oadoi on Nov 02 2018.

\bibitem{galenson2014codehint}
Joel Galenson, Philip Reames, Rastislav Bodik, Bj{\"o}rn Hartmann, and Koushik
  Sen.
\newblock Codehint: Dynamic and interactive synthesis of code snippets.
\newblock In {\em Proceedings of the 36th international conference on Software
  Engineering}, pages 653--663, 2014.

\bibitem{gao2020feedback}
Xiang Gao, Shraddha Barke, Arjun Radhakrishna, Gustavo Soares, Sumit Gulwani,
  Alan Leung, Nachiappan Nagappan, and Ashish Tiwari.
\newblock Feedback-driven semi-supervised synthesis of program transformations.
\newblock {\em Proceedings of the ACM on Programming Languages},
  4(OOPSLA):1--30, 2020.

\bibitem{gelman2013bayesian}
Andrew Gelman, John~B Carlin, Hal~S Stern, David~B Dunson, Aki Vehtari, and
  Donald~B Rubin.
\newblock {\em Bayesian data analysis}.
\newblock CRC press, 2013.

\bibitem{gu2016deep}
Xiaodong Gu, Hongyu Zhang, Dongmei Zhang, and Sunghun Kim.
\newblock Deep api learning.
\newblock In {\em Proceedings of the 2016 24th ACM SIGSOFT International
  Symposium on Foundations of Software Engineering}, pages 631--642, 2016.

\bibitem{deckard}
Lingxiao Jiang, Ghassan Misherghi, Zhendong Su, and Stephane Glondu.
\newblock Deckard: Scalable and accurate tree-based detection of code clones.
\newblock In {\em Proceedings of the 29th International Conference on Software
  Engineering}, ICSE '07, pages 96--105, Washington, DC, USA, 2007. IEEE
  Computer Society.

\bibitem{kamiya2002ccfinder}
Toshihiro Kamiya, Shinji Kusumoto, and Katsuro Inoue.
\newblock Ccfinder: A multilinguistic token-based code clone detection system
  for large scale source code.
\newblock {\em IEEE Transactions on Software Engineering}, 28(7):654--670,
  2002.

\bibitem{koschke2006clone}
Rainer Koschke, Raimar Falke, and Pierre Frenzel.
\newblock Clone detection using abstract syntax suffix trees.
\newblock In {\em 2006 13th Working Conference on Reverse Engineering}, pages
  253--262. IEEE, 2006.

\bibitem{li2004cp}
Zhenmin Li, Shan Lu, Suvda Myagmar, and Yuanyuan Zhou.
\newblock Cp-miner: A tool for finding copy-paste and related bugs in operating
  system code.
\newblock In {\em OSdi}, volume~4, pages 289--302, 2004.

\bibitem{liang2010type}
Percy Liang, Michael~I Jordan, and Dan Klein.
\newblock Type-based mcmc.
\newblock In {\em Human Language Technologies: The 2010 Annual Conference of
  the North American Chapter of the Association for Computational Linguistics},
  pages 573--581, 2010.

\bibitem{mandelin2005jungloid}
David Mandelin, Lin Xu, Rastislav Bod{\'\i}k, and Doug Kimelman.
\newblock Jungloid mining: helping to navigate the api jungle.
\newblock {\em ACM Sigplan Notices}, 40(6):48--61, 2005.

\bibitem{meng2015does}
Na~Meng, Lisa Hua, Miryung Kim, and Kathryn~S McKinley.
\newblock Does automated refactoring obviate systematic editing?
\newblock In {\em 2015 IEEE/ACM 37th IEEE International Conference on Software
  Engineering}, volume~1, pages 392--402. IEEE, 2015.

\bibitem{meng2011systematic}
Na~Meng, Miryung Kim, and Kathryn~S McKinley.
\newblock Systematic editing: generating program transformations from an
  example.
\newblock {\em ACM SIGPLAN Notices}, 46(6):329--342, 2011.

\bibitem{meng2013lase}
Na~Meng, Miryung Kim, and Kathryn~S McKinley.
\newblock Lase: locating and applying systematic edits by learning from
  examples.
\newblock In {\em 2013 35th International Conference on Software Engineering
  (ICSE)}, pages 502--511. IEEE, 2013.

\bibitem{miltner2019fly}
Anders Miltner, Sumit Gulwani, Vu~Le, Alan Leung, Arjun Radhakrishna, Gustavo
  Soares, Ashish Tiwari, and Abhishek Udupa.
\newblock On the fly synthesis of edit suggestions.
\newblock {\em Proceedings of the ACM on Programming Languages},
  3(OOPSLA):1--29, 2019.

\bibitem{mou2016convolutional}
Lili Mou, Ge~Li, Lu~Zhang, Tao Wang, and Zhi Jin.
\newblock Convolutional neural networks over tree structures for programming
  language processing.
\newblock In {\em Proceedings of the AAAI Conference on Artificial
  Intelligence}, volume~30, 2016.

\bibitem{murphy2012machine}
Kevin~P Murphy.
\newblock {\em Machine learning: a probabilistic perspective}.
\newblock MIT press, 2012.

\bibitem{nguyen2009graph}
Tung~Thanh Nguyen, Hoan~Anh Nguyen, Nam~H Pham, Jafar~M Al-Kofahi, and Tien~N
  Nguyen.
\newblock Graph-based mining of multiple object usage patterns.
\newblock In {\em Proceedings of the 7th joint meeting of the European Software
  Engineering Conference and the ACM SIGSOFT symposium on the Foundations of
  Software Engineering}, pages 383--392, 2009.

\bibitem{polozov2015flashmeta}
Oleksandr Polozov and Sumit Gulwani.
\newblock Flashmeta: A framework for inductive program synthesis.
\newblock In {\em Proceedings of the 2015 ACM SIGPLAN International Conference
  on Object-Oriented Programming, Systems, Languages, and Applications}, pages
  107--126, 2015.

\bibitem{premtoon2020semantic}
Varot Premtoon, James Koppel, and Armando Solar-Lezama.
\newblock Semantic code search via equational reasoning.
\newblock In {\em PLDI}, pages 1066--1082, 2020.

\bibitem{rattan2013software}
Dhavleesh Rattan, Rajesh Bhatia, and Maninder Singh.
\newblock Software clone detection: A systematic review.
\newblock {\em Information and Software Technology}, 55(7):1165--1199, 2013.

\bibitem{reiss2009semantics}
Steven~P Reiss.
\newblock Semantics-based code search.
\newblock In {\em 2009 IEEE 31st International Conference on Software
  Engineering}, pages 243--253. IEEE, 2009.

\bibitem{rich1988programmer}
Charles Rich and Richard~C. Waters.
\newblock The programmer's apprentice: A research overview.
\newblock {\em Computer}, 21(11):10--25, 1988.

\bibitem{rolim2017learning}
Reudismam Rolim, Gustavo Soares, Loris D'Antoni, Oleksandr Polozov, Sumit
  Gulwani, Rohit Gheyi, Ryo Suzuki, and Bj{\"o}rn Hartmann.
\newblock Learning syntactic program transformations from examples.
\newblock In {\em 2017 IEEE/ACM 39th International Conference on Software
  Engineering (ICSE)}, pages 404--415. IEEE, 2017.

\bibitem{rolim2018learning}
Reudismam Rolim, Gustavo Soares, Rohit Gheyi, Titus Barik, and Loris D'Antoni.
\newblock Learning quick fixes from code repositories.
\newblock {\em arXiv preprint arXiv:1803.03806}, 2018.

\bibitem{rosen1988global}
Barry~K Rosen, Mark~N Wegman, and F~Kenneth Zadeck.
\newblock Global value numbers and redundant computations.
\newblock In {\em Proceedings of the 15th ACM SIGPLAN-SIGACT symposium on
  Principles of programming languages}, pages 12--27, 1988.

\bibitem{roy2008nicad}
Chanchal~K Roy and James~R Cordy.
\newblock Nicad: Accurate detection of near-miss intentional clones using
  flexible pretty-printing and code normalization.
\newblock In {\em Program Comprehension, 2008. ICPC 2008. The 16th IEEE
  International Conference on}, pages 172--181. IEEE, 2008.

\bibitem{sajnani2016sourcerercc}
Hitesh Sajnani, Vaibhav Saini, Jeffrey Svajlenko, Chanchal~K Roy, and
  Cristina~V Lopes.
\newblock Sourcerercc: Scaling code clone detection to big-code.
\newblock In {\em Software Engineering (ICSE), 2016 IEEE/ACM 38th International
  Conference on}, pages 1157--1168. IEEE, 2016.

\bibitem{sivaraman2019active}
Aishwarya Sivaraman, Tianyi Zhang, Guy Van~den Broeck, and Miryung Kim.
\newblock Active inductive logic programming for code search.
\newblock In {\em 2019 IEEE/ACM 41st International Conference on Software
  Engineering (ICSE)}, pages 292--303. IEEE, 2019.

\bibitem{stolee2014solving}
Kathryn~T Stolee, Sebastian Elbaum, and Daniel Dobos.
\newblock Solving the search for source code.
\newblock {\em ACM Transactions on Software Engineering and Methodology
  (TOSEM)}, 23(3):1--45, 2014.

\bibitem{wang2013mining}
Jue Wang, Yingnong Dang, Hongyu Zhang, Kai Chen, Tao Xie, and Dongmei Zhang.
\newblock Mining succinct and high-coverage api usage patterns from source
  code.
\newblock In {\em 2013 10th Working Conference on Mining Software Repositories
  (MSR)}, pages 319--328. IEEE, 2013.

\bibitem{wang2020detecting}
Wenhan Wang, Ge~Li, Bo~Ma, Xin Xia, and Zhi Jin.
\newblock Detecting code clones with graph neural network and flow-augmented
  abstract syntax tree.
\newblock In {\em 2020 IEEE 27th International Conference on Software Analysis,
  Evolution and Reengineering (SANER)}, pages 261--271. IEEE, 2020.

\bibitem{wang2010matching}
Xiaoyin Wang, David Lo, Jiefeng Cheng, Lu~Zhang, Hong Mei, and Jeffrey~Xu Yu.
\newblock Matching dependence-related queries in the system dependence graph.
\newblock In {\em Proceedings of the IEEE/ACM international conference on
  Automated software engineering}, pages 457--466, 2010.

\bibitem{zhong2009mapo}
Hao Zhong, Tao Xie, Lu~Zhang, Jian Pei, and Hong Mei.
\newblock Mapo: Mining and recommending api usage patterns.
\newblock In {\em European Conference on Object-Oriented Programming}, pages
  318--343. Springer, 2009.

\end{thebibliography}
\end{document}